\documentclass[prl,letterpaper,aps,floatfix,twocolumn]{revtex4-1}
\usepackage{graphicx}
\usepackage{amsmath}
\usepackage{subfigure}
\newcommand{\beq}{\begin{equation}}
\newcommand{\eeq}{\end{equation}}

\newcommand{\bB}{{\bf{B}}}

\newcommand{\bj}{{\bf j}}
\newcommand{\beqa}{\begin{eqnarray}}
\newcommand{\eeqa}{\end{eqnarray}}

\newcommand{\bnabla}{{\boldsymbol \nabla}}

\begin{document}
\title{Giant Planar Hall Effect in Topological Metals}
\author{A.A. Burkov}
\affiliation{Department of Physics and Astronomy, University of Waterloo, Waterloo, Ontario 
N2L 3G1, Canada} 
\date{\today}
\begin{abstract}
Much excitement has been generated recently by the experimental observation of 
the chiral anomaly in condensed matter physics. This manifests as strong negative longitudinal 
magnetoresistance and has so far been clearly observed in Na$_3$Bi, ZrTe$_5$, and GdPtBi. 
In this work we point out that the chiral anomaly must lead to another effect in topological metals, that has 
been overlooked so far: Giant Planar Hall Effect (GPHE), which is the appearance of a large transverse voltage when the 
in plane magnetic field is not aligned with the current. 
Moreover, we demonstrate that the GPHE is closely related to the angular narrowing of the negative longitudinal magnetoresistance signal, observed experimentally. 
\end{abstract}
\maketitle
The recent theoretical~\cite{Wan11,Ran11,Burkov11-1,Burkov11-2,Xu11,Kane12,Fang12,Fang13}
and experimental~\cite{Chen14,Neupane14,HasanTaAs,DingTaAs2,DingTaAs,Lu15}
discovery of Dirac and Weyl semimetals has extended the notions of nontrivial electronic structure topology to metals. 
It has also reinforced the connection that exists between the physics of materials with topologically nontrivial 
electronic structure and the physics of relativistic fermions. 
Chiral anomaly, which refers to nonconservation of the chiral charge in the presence of collinear external electric and magnetic fields, is a particularly important example that highlights such a connection. 
Discovered theoretically by Adler~\cite{Adler69} and by Bell and Jackiw~\cite{Jackiw69} in the relativistic particle physics context, 
it provided the explanation 
for the observed fast decay of a neutral pion into two photons, naively not allowed by the chiral charge conservation. 
Very recently, a condensed matter manifestation of the chiral anomaly was finally observed in Dirac semimetals 
Na$_3$Bi~\cite{Ong_anomaly}, ZrTe$_5$~\cite{Li_anomaly}, and in a half-Heusler compound GdPtBi~\cite{Ong_GdPtBi}.

The chiral anomaly manifests in Weyl and Dirac semimetals as a very unusual large negative longitudinal magnetoresistance, quadratic in the applied magnetic field, as predicted theoretically~\cite{Spivak12,Burkov_lmr_prl,Burkov_lmr_prb}. 
While the existence of the effect, and most of its observed features are in qualitative agreement with the theory, one puzzling feature
has remained unexplained. As was first pointed out in~\cite{Ong_anomaly}, the observed dependence of the magnetoresistance 
on the angle $\theta$ between the current and the applied magnetic field, is against the expectations, drawn from the existing theory. 
Namely, the theory of Refs.~\cite{Spivak12,Burkov_lmr_prl,Burkov_lmr_prb} naively predicts a $\cos^2 \theta$ dependence, due to the quadratic dependence of the chiral anomaly contribution to the conductivity on the magnetic field, but the observed angular dependence appears to be much stronger.  

In this paper we both explain the angular narrowing phenomenon and connect it with another effect, which has 
so far not been mentioned in relation to the chiral anomaly, the Giant Planar Hall Effect (GPHE). 
Note that a closely related explanation of the angular narrowing effect has already been proposed by us in Ref.~\cite{Burkov16}, 
but the connection with the GPHE was not understood there. 
We argue that the presence of both the negative longitudinal magnetoresistance with a characteristic dependence on the 
angle between the current and the magnetic field, and the GPHE may be regarded as a smoking gun 
signature of the chiral anomaly. 

As was argued in Refs.~\cite{Spivak12,Parameswaran14,Burkov_lmr_prl,Burkov_lmr_prb}, transport in topological (both Weyl and Dirac) metals 
is distinguished by the existence of an extra (nearly) conserved quantity, the chiral charge, which is coupled to the electric charge 
in the presence of an external magnetic field. 
The hydrodynamic transport equations for the electric and the chiral charge have the following 
form~\cite{Burkov_lmr_prl,Burkov_lmr_prb}
\beqa
\label{eq:1}
\frac{\partial n}{\partial t}&=&D \bnabla^2 (n + g V) + \Gamma \bB \cdot \bnabla (n_c + g V_c), \nonumber \\
\frac{\partial n_c}{\partial t}&=&D \bnabla^2 (n_c + g V_c) - \frac{n_c + g V_c}{\tau_c} + \Gamma \bB \cdot \bnabla (n + g V). 
\nonumber \\
\eeqa
Here $-e n$ is the electric charge density, $-e n_c$ is the chiral charge density (defined as the difference between the total right handed and total left handed charge); $D$ is the diffusion coefficient (we take the diffusion coefficients, corresponding to the electric and the chiral charges to be the same for simplicity, although they may in general be different due to electron-electron interaction effects); $g$ is the density of states at the Fermi energy;
$\Gamma = e/ 2 \pi^2 g$ is a transport coefficient, which characterizes the chiral anomaly induced coupling between the 
electric and the chiral charge in the presence of an applied magnetic field $\bB$; and $\tau_c$ is the chiral charge relaxation time, 
which is taken to be long, reflecting the near conservation of the chiral charge. 
We will use $\hbar = c = 1$ units throughout this paper, except in some of the final results. 

The presence of the electrostatic potential $V$ and the ``chiral electrostatic potential" $V_c$ (this is a hypothetical external potential that couples antisymmetrically to the right- and left-handed charge) reflects the presence of both diffusion and drift contributions to the electric and chiral currents correspondingly. In equilibrium the two contributions must cancel each other, which, in particular, implies $n_c + g V_c = 0$ in this case.  We note that a time-independent spatially-uniform $V_c$ will always be present in a noncentrosymmetric topological metal~\cite{Zyuzin12-2}.
\begin{figure}[t]
\vspace{-2cm}
\hspace{-1.5cm}
\includegraphics[width=10cm]{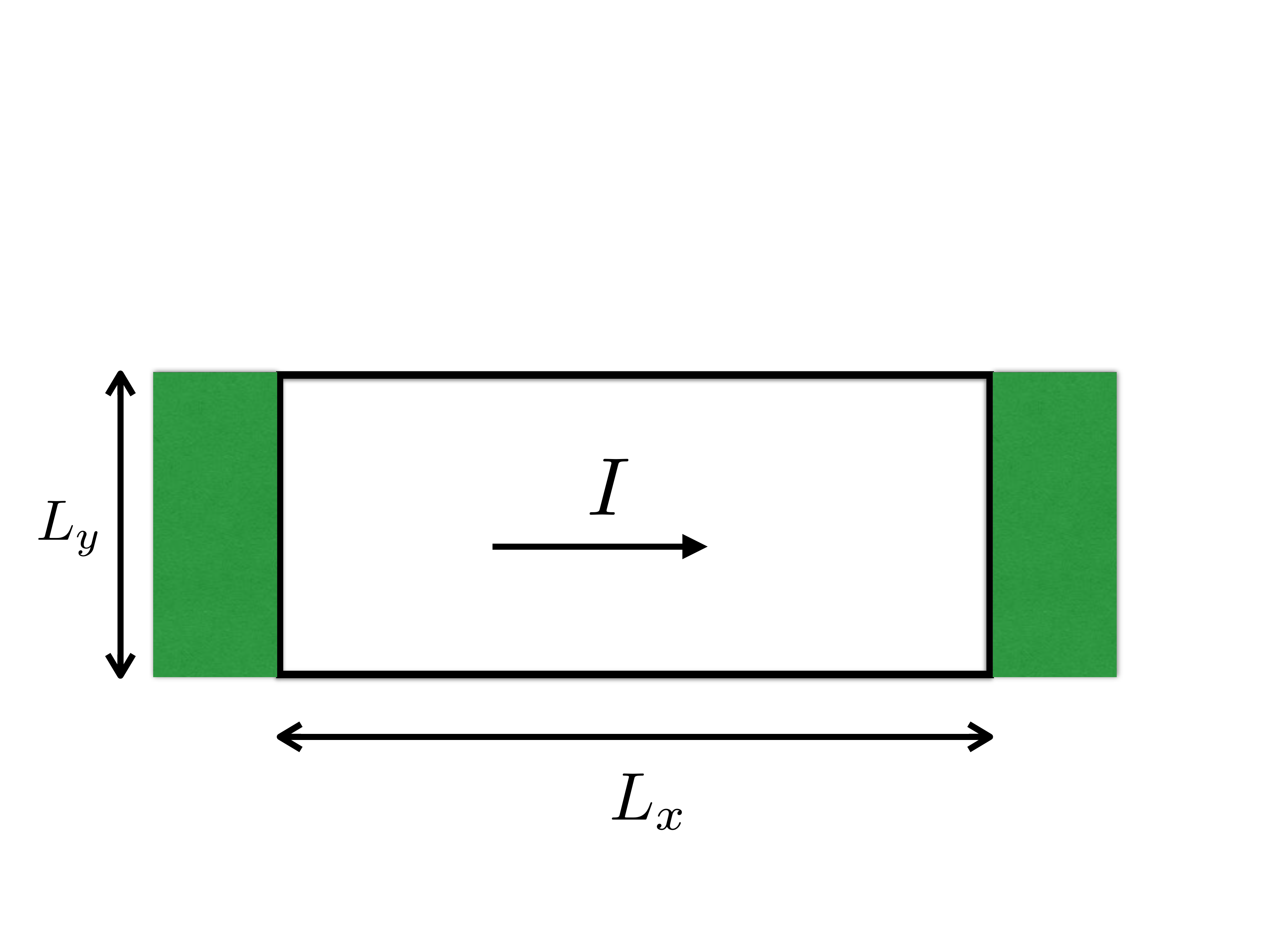}
\vspace{-1cm}
\caption{(Color online) Schematics of the sample setup. Ordinary metal electrodes (green) are attached along the whole width of the sample cross-section and inject current into the sample uniformly. The sample has a square cross-section of area $L_y^2$ and length $L_x$.}
\label{fig:1}
\end{figure}

Let us consider an experimental setup, shown in Fig.~\ref{fig:1}. 
We will assume a sample of length $L_x$ in the $x$ direction, attached to current-carrying normal (i.e. non-topological) metallic
leads at $x = \pm L_x/2$, and a square (for simplicity) cross-section of area $L_y^2$. 
Suppose electric current $I$ is injected and extracted uniformly at the attached leads.
We want to find the voltage that develops in response to this current, or the resistivity tensor of our system. 

It is convenient to introduce ``electrochemical potentials" $\mu = (n + g V)/g$ and $\mu_c = (n_c + g V_c)/g$. 
The electric current density is then given by
\beq
\label{eq:2}
\bj = \frac{\sigma}{e} \bnabla \mu + e g \Gamma \mu_c \bB, 
\eeq
where $\sigma = e^2 g D$ is the Drude conductivity.
The second term in Eq.~\eqref{eq:2} expresses the chiral magnetic effect (CME)~\cite{Kharzeev08}. 
Since $\mu_c = 0$ in equilibrium, CME vanishes in equilibrium as it should~\cite{Franz13,Chen13}. 

We assume that in the steady state the current only flows in the $x$ direction, which means 
\beq
\label{eq:3}
j_x = \frac{I}{L_y^2}, \,\, j_y = j_z = 0. 
\eeq
We will also assume that the magnetic field is only rotated in the $xy$ plane ($xz$ plane is identical). 
This implies that we may take both electrochemical potentials $\mu$ and $\mu_c$ to be independent of $z$. 
Then Eqs.~\eqref{eq:2},\eqref{eq:3} allow us to express $\bnabla \mu$ in terms of the current and the chiral electrochemical 
potential $\mu_c$ as
\beqa
\label{eq:4}
\frac{\partial \mu}{\partial x}&=&\frac{e I}{\sigma L_y^2} - \frac{\mu_c}{L_a} \cos \theta, \nonumber \\
\frac{\partial \mu}{\partial y}&=&- \frac{\mu_c}{L_a} \sin \theta, 
\eeqa
where $\theta = \arctan(B_y/B_x)$ is the angle between he applied magnetic field and the current and 
we have introduced a magnetic field related length scale
\beq
\label{eq:5}
L_a = \frac{D}{\Gamma B}, 
\eeq
which will play a crucial role in what follows. Note that this length scale is distinct from the usual magnetic length 
$\ell_B = 1/\sqrt{eB}$ and 
appears due to the chiral anomaly (hence the subscript a). 
$1/L_a$ quantifies the strength of the chiral anomaly induced coupling between the electric and the chiral charge. 
The existence of this new length scale was first pointed out in Ref.~\cite{Altland15}. 
Substituting Eq.~\eqref{eq:4} into the equation for the chiral electrochemical potential, we obtain
\beq
\label{eq:6}
\frac{\partial^2 \mu_c}{\partial x^2} + \frac{\partial^2 \mu_c}{\partial y^2} - \frac{\mu_c}{\lambda^2} = - 
\frac{e I \cos \theta}{\sigma L_a L_y^2},
\eeq
where 
\beq
\label{eq:7}
\lambda^2 = \frac{L_a^2 L_c^2}{L_a^2 + L_c^2}, 
\eeq
and we have introduced another important length scale $L_c = \sqrt{D \tau_c}$, which has the meaning of the chiral 
charge diffusion length.
Transport effects due to the chiral anomaly may be expected to be significant only when $L_c$ is a macroscopic length 
scale, i.e. when the chiral charge is a nearly conserved quantity. 
More specifically, as will be shown below, the parameter that determines the strength of the chiral anomaly induced
magnetotransport effects in topological metals is the ratio of the two length scales $L_c/L_a$. 

We may further simplify Eq.~\eqref{eq:6} by noticing that the condition of no charge current in the $y$ direction, 
expressed by the second of Eqs.~\eqref{eq:4}, may always be satisfied by taking $\mu_c$ to be independent of $y$, which 
then implies uniform gradient of the electrochemical potential $\mu$ in the $y$ direction. 
Eq.~\eqref{eq:6} then simplifies to 
\beq
\label{eq:8}
\frac{d^2 \mu_c}{d x^2} - \frac{\mu_c}{\lambda^2} = - \frac{e I \cos \theta}{\sigma L_a L_y^2}.
\eeq

Equation~\eqref{eq:8} needs to be solved with the appropriate boundary conditions.
These are naturally not universal and depend on the details of the experimental setup being modelled. 
We will assume that the sample is attached to uniform current carrying leads across the whole width of the sample cross-section, 
and the lead material is a normal non-topological metal. In this case, chiral charge must rapidly relax upon entering the normal 
leads and the most appropriate boundary condition is thus of the Dirichlet type
\beq
\label{eq:9}
\mu_c(x = \pm L_x/2) = 0. 
\eeq
We note, however, that the boundary conditions do not affect the final results at all for large sample sizes $L_x \gg L_a, L_c$
(scale dependence of transport coefficients is, however, interesting in its own right in this case and may be observable due to the 
large size of $L_c$~\cite{Altland15}). 

Solving Eq.~\eqref{eq:8} with the above boundary conditions, we obtain
\beq
\label{eq:10}
\mu_c(x) = \frac{e I \lambda^2 \cos \theta}{\sigma L_a L_y^2} \left[1 - \frac{\cosh(x/\lambda)}{\cosh(L_x/2 \lambda)}\right]. 
\eeq
Substituting Eq.~\eqref{eq:10} into Eq.~\eqref{eq:4}, we may now calculate the voltage drops that develop across the sample 
in the $x$ and $y$ directions in response to the current in the $x$ direction, as integrals of the corresponding 
electrochemical potential gradients. We obtain
\beqa
\label{eq:11}
V_x&=&\frac{1}{e} \int_{-L_x/2}^{L_x/2}\, d x\, \frac{\partial \mu}{\partial x} = \frac{I L_x}{\sigma L_y^2} 
\left(1 - \frac{\lambda^2}{L_a^2} \cos^2 \theta \right) \nonumber \\
&+&\frac{2 I \lambda^3 \cos^2 \theta }{\sigma L_y^2 L_a^2} \tanh(L_x/2 \lambda), 
\eeqa
and 
\beqa
\label{eq:12}
V_y&=&\frac{1}{e L_x} \int_{-L_x/2}^{L_x/2}\, d x\, \int_{-L_y/2}^{L_y/2}\, d y\, \frac{\partial \mu}{\partial y} \nonumber \\
&=&- \frac{I \lambda^2 \cos \theta \sin \theta}{\sigma L_a^2 L_y} \left[1 - \frac{2 \lambda}{L_x} \tanh(L_x/2 \lambda) \right],
\eeqa
where we have averaged $V_y$ along the length of the sample in the $x$ direction. 

Eqs.~\eqref{eq:11} and \eqref{eq:12} then imply the following result for the scale-dependent resistivity tensor 
\beqa
\label{eq:13}
\rho_{xx}&=&\frac{1}{\sigma}\left(1 - \frac{\lambda^2}{L_a^2} \cos^2 \theta\right) + 
\frac{2 \lambda^2 \cos^2 \theta}{\sigma L_a^2 L_x} \tanh(L_x/2 \lambda), \nonumber \\
\rho_{yx}&=&- \frac{\lambda^2 \sin \theta \cos \theta}{\sigma L_a^2} \left[1 - \frac{2 \lambda}{L_x} \tanh(L_x/ 2 \lambda) \right]. 
\eeqa
The resistivity tensor thus contains both diagonal and off-diagonal components, the off-diagonal ones being induced 
by the magnetic field. 
Since $\lambda$ is always dominated by the shortest of the two length scales $L_{a,c}$, it is clear that the off-diagonal 
resistivity vanishes when $L_a$ diverges, making it clear that the origin of the off-diagonal component of the resistivity 
tensor is the chiral anomaly. 
More specifically, its origin can be easily traced back to the CME contribution to the electrical current in Eq.~\eqref{eq:2}. 

Eq.~\eqref{eq:13} may be rewritten in a more illuminating form in terms of $\rho_{\parallel}$ and $\rho_{\perp}$, 
i.e. diagonal components of the resistivity tensor, corresponding to the current flow along and perpendicular 
to the direction of the magnetic field. 
From Eq.~\eqref{eq:13}, we have
\beq
\label{eq:14}
\rho_{\parallel} = \frac{1}{\sigma} \left(1 - \frac{\lambda^2}{L_a^2} \right) + \frac{2 \lambda^3}{\sigma L_a^2 L_x} 
\tanh(L_x/ 2 \lambda), \,\, \rho_{\perp} = \frac{1}{\sigma}.
\eeq
Then Eq.~\eqref{eq:13} may be written as
\beqa
\label{eq:15}
\rho_{xx}&=&\rho_{\perp} - \Delta \rho \cos^2 \theta, \nonumber \\
\rho_{yx}&=&- \Delta \rho \sin \theta \cos \theta, 
\eeqa
where $\Delta \rho = \rho_{\perp} - \rho_{\parallel}$ is the chiral anomaly induced resistivity anisotropy. 

Eq.~\eqref{eq:15} has the form of a standard relation between the anisotropic magnetoresistance (AMR), represented by the first equation, and the planar Hall effect (PHE), which is expressed by the second equation~\cite{Pan57,Giordano95,Awschalom_gphe}. 
Note that the name PHE is a bit of a misnomer: the off-diagonal 
resistivity does not satisfy the antisymmetry property of a true Hall effect $\rho_{xy} = - \rho_{yx}$, since it does not 
originate from the Lorentz force. 
It is, however, the standard name for this phenomenon in the literature and we will thus use it as well. 
Both AMR and PHE are well known phenomena in ferromagnetic metals, originating in this case from the interplay of the magnetic order and the spin-orbit interactions. Both are typically very weak, but 
can be much stronger in ferromagnets with significant spin-orbit interactions, such as doped magnetic 
semiconductors~\cite{Awschalom_gphe}. 
What is remarkable about our result is that neither AMR nor PHE in a topological metal require magnetic order, originating instead 
from the chiral anomaly, and their magnitude can be extremely large (in fact, approaching the theoretical upper limit at increasing magnetic field), as we show below. 
The sign of the effect in our case is also opposite to what is typically observed in ferromagnets: $\Delta \rho$, as defined in 
Eq.~\eqref{eq:15}, is positive in our case, but would typically be negative in a metallic ferromagnet~\cite{Giordano95}. 

The parallel resistivity $\rho_{\parallel}$ exhibits a nontrivial dependence on the two intrinsic length scales $L_{a,c}$ 
of the material and on the sample size $L_x$. 
Let us first consider the regime of weak magnetic fields, corresponding to $L_a \gg L_c$. 
In this case, taking the limit of large sample size $L_x \gg L_c$, we obtain
\beq
\label{eq:16}
\rho_{\parallel} \approx \frac{1}{\sigma}\left[1 - \left(\frac{L_c}{L_a}\right)^2 \right],
\eeq
which corresponds to a small negative quadratic magnetic field dependent correction to the longitudinal resistivity. 
The PHE in this case is also small and given by
\beq
\label{eq:17}
\rho_{yx} = - \frac{1}{\sigma}\left(\frac{L_c}{L_a} \right)^2\sin \theta \cos \theta. 
\eeq

A more interesting regime is the regime of stronger magnetic field, corresponding to $L_a \ll L_c$. 
Assuming the sample size $L_x > L_a$, we obtain
\beq
\label{eq:18}
\rho_{\parallel} = \frac{L_a^2}{\sigma L_c^2} \left(1 + \frac{2 L_c^2}{L_a L_x} \right). 
\eeq
Eq.~\eqref{eq:18} exhibits an interesting and nontrivial scale dependence. 
Indeed, suppose that $L_a < L_x < L_c^2/L_a$ (the upper limit is a third nontrivial length scale in this problem!). 
In this case 
\beq
\label{eq:19}
\rho_{\parallel} \approx \frac{2 L_a}{\sigma L_x} = \frac{4 \pi^2 \ell_B^2}{e^2 L_x}. 
\eeq
To understand the meaning of this result it is convenient to evaluate the corresponding conductance
\beq
\label{eq:20}
G_{\parallel} = \frac{\rho_{\parallel} L_y^2}{L_x} = \frac{e^2 N_{\phi}}{2 \pi}, 
\eeq
where $N_{\phi} = L_y^2/2 \pi \ell_B^2$ is the number of magnetic flux quanta, penetrating the sample cross section. 
This is identical to the result of Ref.~\cite{Altland15}, obtained by a different method. 
Physically, this corresponds to a regime, in which the sample conductance is dominated by the chiral lowest Landau level, 
which is where the chiral anomaly contribution to Eqs.~\eqref{eq:1}, \eqref{eq:2} comes from~\cite{Burkov_lmr_prl,Burkov_lmr_prb}. 
$G_{\parallel}$ then corresponds to a conductance of $e^2/h$ per lowest Landau level orbital state, i.e. is identical 
to the conductance of an effective one-dimensional system with $N_{\phi}$ conduction channels. 

Most importantly, the resistivity anisotropy and thus the magnitude of the PHE in this regime is given by 
\beq
\label{eq:21}
\Delta \rho = \frac{1}{\sigma}\left(1 - \frac{2 L_a}{L_x}\right). 
\eeq
Thus $\Delta \rho$ is starting to approach its maximal possible value of $1/\sigma$ when the sample 
size is increased.
We thus call this Giant Planar Hall Effect (GPHE) (this name was first used in relation to PHE in the context of magnetic semiconductors in Ref.~\cite{Awschalom_gphe}). 

For larger sample sizes, when $L_x > L_c^2/L_a$, the magnetic field dependence of the longitudinal 
resistivity crosses over from $1/B$ to $1/B^2$
\beq
\label{eq:22}
\rho_{\parallel} \approx \frac{1}{\sigma} \left(\frac{L_a}{L_c}\right)^2. 
\eeq
The GPHE magnitude in this case becomes independent of the sample size and has reached its maximal magnitude
\beq
\label{eq:23}
\Delta \rho = \frac{1}{\sigma}\frac{(L_c/L_a)^2}{1 + (L_c/L_a)^2},
\eeq
which converges to $1/\sigma$ as the magnetic field is increased. 

Interestingly, there exists a direct connection between the GPHE and the angular narrowing of the negative longitudinal magnetoconductivity signal~\cite{Ong_anomaly}, as we will now demonstrate.  
Using Eq.~\eqref{eq:15}, we obtain
\beqa
\label{eq:24}
\rho^{-1}_{xx}(B) - \rho^{-1}_{xx}(0)&=&\frac{\sigma (\Delta \rho/\rho_{\parallel}) \cos^2 \theta}
{1 + (\Delta \rho/ \rho_{\parallel})\sin^2 \theta} \nonumber \\
&=&\frac{\sigma (L_c/L_a)^2 \cos^2 \theta}{1 + (L_c/L_a)^2\sin^2 \theta}.
\eeqa
What is missing in the standard expressions for the chiral anomaly induced magnetoconductivity~\cite{Spivak12} is the 
angular dependence in the denominator in Eq.~\eqref{eq:24}. 
This clearly leads to narrowing of the angular dependence: Eq.~\eqref{eq:24} at small angles has the form of a Lorentzian 
with the angular width $\Delta \theta \sim L_a/ L_c$,
which gets narrower as the magnitude of the chiral anomaly induced GPHE increases.
This nontrivial connection between the GPHE and the angular narrowing of the negative longitudinal magnetoresistance 
signal may be regarded as smoking gun evidence for the chiral anomaly. 
\begin{figure}[t]
\includegraphics[width=8cm]{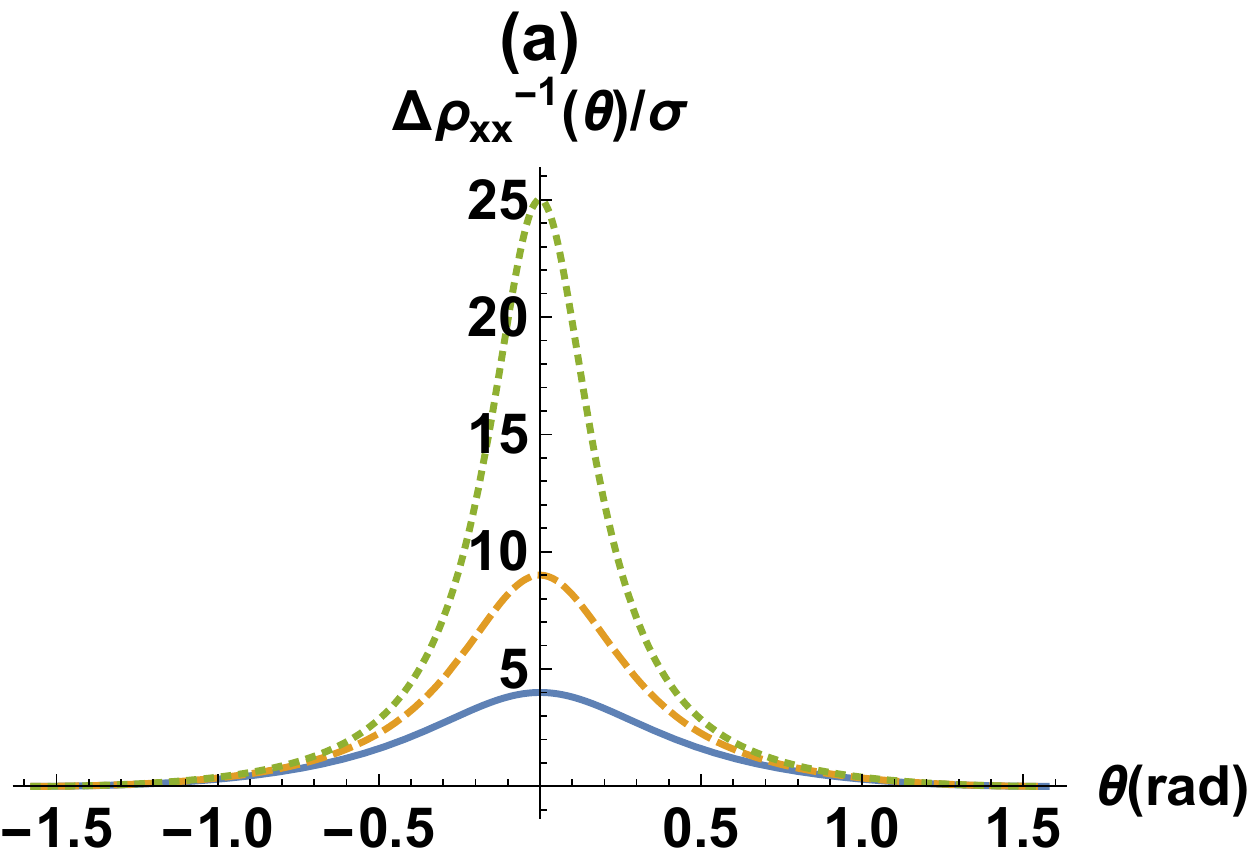}
\includegraphics[width=8cm]{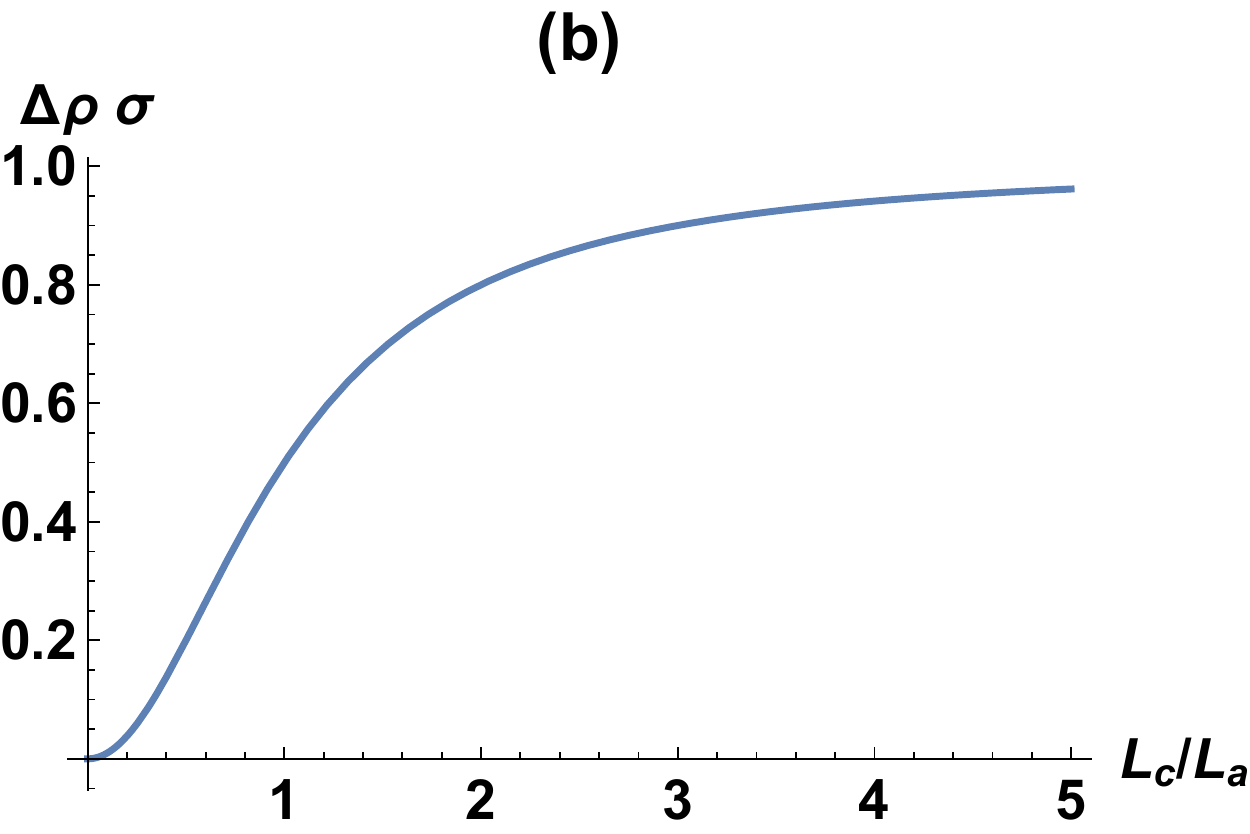}
\caption{(Color online) (a) Inverse longitudinal magnetoresistivity as a function of the angle between the current and the magnetic field. Different curves correspond to different values of the ratio $L_c/L_a$: 2 (blue, solid), 3 (orange, dashed) and  
5 (green, dotted). (b) Dependence of the resistivity anisotropy $\Delta \rho$ and thus the magnitude of the GPHE on $L_c/L_a$.}
\label{fig:2}
\end{figure}

In order to relate our results to the existing experimental data~\cite{Ong_anomaly,Li_anomaly}, 
it is useful to express the ratio $L_c/L_a$ explicitly in terms of the magnetic field. 
We obtain
\beq
\label{eq:25}
\frac{L_c}{L_a} = \Gamma B \sqrt{\frac{\tau_c}{D}} \sim \left(\frac{\hbar v_F/ \ell_B}{\epsilon_F}\right)^2 \sqrt{\frac{\tau_c}{\tau}}, 
\eeq
where $v_F$ is the Fermi velocity of the Weyl (Dirac) fermions, $\epsilon_F$ is the Fermi energy and $\tau$ is the momentum relaxation time. 
Taking the values for these parameters from Ref.~\cite{Ong_anomaly}, we have 
$v_F \approx 3.5 \cdot 10^{7}\textrm{cm/s}$, $\epsilon_F \approx 30 \,\textrm{meV}$, and $\tau_c/\tau \approx 50$. 
Assuming the density of states to be $g = \epsilon_F^2/2 \pi^2 \hbar^3 v_F^3$, and 
substituting the above values in Eq.~\eqref{eq:25}, we obtain $L_c/L_a \approx 0.7 \, B/ 1 \, \textrm{Tesla}$. 

In Fig.~\ref{fig:2} we plot both the angular dependence of the inverse longitudinal magnetoresistivity 
and the magnitude of the GPHE using $L_c/L_a$ values, which correspond to the range of magnetic fields of up to about 
$10\, \textrm{T}$, which was used in Refs.~\cite{Ong_anomaly,Li_anomaly}. 
Qualitatively, the behavior of the magnetoresistivity appears to agree with the experimental data. 
In particular, at low magnetic fields the magnetoconductivity peak follows a $B^2$ dependence, while the 
angular width goes roughly as $1/B$, consistent with Eqs.~\eqref{eq:24} and \eqref{eq:25}. 
At higher fields both appear to saturate, which is likely explained by the fact that in both experiments the quantum regime
$\epsilon_F < \hbar v_F/ \ell_B$ is reached at magnetic fields of just a few Tesla. 
In this regime we expect $g \sim B$ and $1/\tau \sim B$, $1/\tau_c \sim B$, thus making the ratio $L_c/L_a$ independent 
of the magnetic field. 
Fig.~\ref{fig:2} also shows that the magnitude of the GPHE may approach its maximal value of $\Delta \rho = 1/\sigma$ for 
experimentally accessible values of the magnetic field. 

In conclusion, we have described a magnetotransport effect, related to the chiral anomaly, that has not been noticed 
before, the Giant Planar Hall Effect. 
We have also connected this effect to the angular dependence of the longitudinal magnetoconductivity, explaining its magnetic 
field dependent narrowing, pointed out in Ref.~\cite{Ong_anomaly}. 
Observation of both the negative longitudinal magnetoresistance and the GPHE, with a specific relation between 
the angular dependence of the magnetoresistance and the GPHE magnitude, given by Eq.~\eqref{eq:24}, 
constitutes a smoking gun evidence for the chiral anomaly. 
We note that the GPHE may have already been observed in a recent experiment on ZrTe$_5$~\cite{Ong16}. 

\begin{acknowledgments}
We acknowledge useful discussions with Nai Phuan Ong. Financial support was provided by NSERC of Canada. 
\end{acknowledgments}
\bibliography{references}

\begin{thebibliography}{33}%
\makeatletter
\providecommand \@ifxundefined [1]{%
 \@ifx{#1\undefined}
}%
\providecommand \@ifnum [1]{%
 \ifnum #1\expandafter \@firstoftwo
 \else \expandafter \@secondoftwo
 \fi
}%
\providecommand \@ifx [1]{%
 \ifx #1\expandafter \@firstoftwo
 \else \expandafter \@secondoftwo
 \fi
}%
\providecommand \natexlab [1]{#1}%
\providecommand \enquote  [1]{``#1''}%
\providecommand \bibnamefont  [1]{#1}%
\providecommand \bibfnamefont [1]{#1}%
\providecommand \citenamefont [1]{#1}%
\providecommand \href@noop [0]{\@secondoftwo}%
\providecommand \href [0]{\begingroup \@sanitize@url \@href}%
\providecommand \@href[1]{\@@startlink{#1}\@@href}%
\providecommand \@@href[1]{\endgroup#1\@@endlink}%
\providecommand \@sanitize@url [0]{\catcode `\\12\catcode `\$12\catcode
  `\&12\catcode `\#12\catcode `\^12\catcode `\_12\catcode `\%12\relax}%
\providecommand \@@startlink[1]{}%
\providecommand \@@endlink[0]{}%
\providecommand \url  [0]{\begingroup\@sanitize@url \@url }%
\providecommand \@url [1]{\endgroup\@href {#1}{\urlprefix }}%
\providecommand \urlprefix  [0]{URL }%
\providecommand \Eprint [0]{\href }%
\providecommand \doibase [0]{http://dx.doi.org/}%
\providecommand \selectlanguage [0]{\@gobble}%
\providecommand \bibinfo  [0]{\@secondoftwo}%
\providecommand \bibfield  [0]{\@secondoftwo}%
\providecommand \translation [1]{[#1]}%
\providecommand \BibitemOpen [0]{}%
\providecommand \bibitemStop [0]{}%
\providecommand \bibitemNoStop [0]{.\EOS\space}%
\providecommand \EOS [0]{\spacefactor3000\relax}%
\providecommand \BibitemShut  [1]{\csname bibitem#1\endcsname}%
\let\auto@bib@innerbib\@empty
\bibitem [{\citenamefont {Wan}\ \emph {et~al.}(2011)\citenamefont {Wan},
  \citenamefont {Turner}, \citenamefont {Vishwanath},\ and\ \citenamefont
  {Savrasov}}]{Wan11}%
  \BibitemOpen
  \bibfield  {author} {\bibinfo {author} {\bibfnamefont {X.}~\bibnamefont
  {Wan}}, \bibinfo {author} {\bibfnamefont {A.~M.}\ \bibnamefont {Turner}},
  \bibinfo {author} {\bibfnamefont {A.}~\bibnamefont {Vishwanath}}, \ and\
  \bibinfo {author} {\bibfnamefont {S.~Y.}\ \bibnamefont {Savrasov}},\ }\href
  {\doibase 10.1103/PhysRevB.83.205101} {\bibfield  {journal} {\bibinfo
  {journal} {Phys. Rev. B}\ }\textbf {\bibinfo {volume} {83}},\ \bibinfo
  {pages} {205101} (\bibinfo {year} {2011})}\BibitemShut {NoStop}%
\bibitem [{\citenamefont {Yang}\ \emph {et~al.}(2011)\citenamefont {Yang},
  \citenamefont {Lu},\ and\ \citenamefont {Ran}}]{Ran11}%
  \BibitemOpen
  \bibfield  {author} {\bibinfo {author} {\bibfnamefont {K.-Y.}\ \bibnamefont
  {Yang}}, \bibinfo {author} {\bibfnamefont {Y.-M.}\ \bibnamefont {Lu}}, \ and\
  \bibinfo {author} {\bibfnamefont {Y.}~\bibnamefont {Ran}},\ }\href {\doibase
  10.1103/PhysRevB.84.075129} {\bibfield  {journal} {\bibinfo  {journal} {Phys.
  Rev. B}\ }\textbf {\bibinfo {volume} {84}},\ \bibinfo {pages} {075129}
  (\bibinfo {year} {2011})}\BibitemShut {NoStop}%
\bibitem [{\citenamefont {Burkov}\ and\ \citenamefont
  {Balents}(2011)}]{Burkov11-1}%
  \BibitemOpen
  \bibfield  {author} {\bibinfo {author} {\bibfnamefont {A.~A.}\ \bibnamefont
  {Burkov}}\ and\ \bibinfo {author} {\bibfnamefont {L.}~\bibnamefont
  {Balents}},\ }\href {\doibase 10.1103/PhysRevLett.107.127205} {\bibfield
  {journal} {\bibinfo  {journal} {Phys. Rev. Lett.}\ }\textbf {\bibinfo
  {volume} {107}},\ \bibinfo {pages} {127205} (\bibinfo {year}
  {2011})}\BibitemShut {NoStop}%
\bibitem [{\citenamefont {Burkov}\ \emph {et~al.}(2011)\citenamefont {Burkov},
  \citenamefont {Hook},\ and\ \citenamefont {Balents}}]{Burkov11-2}%
  \BibitemOpen
  \bibfield  {author} {\bibinfo {author} {\bibfnamefont {A.~A.}\ \bibnamefont
  {Burkov}}, \bibinfo {author} {\bibfnamefont {M.~D.}\ \bibnamefont {Hook}}, \
  and\ \bibinfo {author} {\bibfnamefont {L.}~\bibnamefont {Balents}},\ }\href
  {\doibase 10.1103/PhysRevB.84.235126} {\bibfield  {journal} {\bibinfo
  {journal} {Phys. Rev. B}\ }\textbf {\bibinfo {volume} {84}},\ \bibinfo
  {pages} {235126} (\bibinfo {year} {2011})}\BibitemShut {NoStop}%
\bibitem [{\citenamefont {Xu}\ \emph {et~al.}(2011)\citenamefont {Xu},
  \citenamefont {Weng}, \citenamefont {Wang}, \citenamefont {Dai},\ and\
  \citenamefont {Fang}}]{Xu11}%
  \BibitemOpen
  \bibfield  {author} {\bibinfo {author} {\bibfnamefont {G.}~\bibnamefont
  {Xu}}, \bibinfo {author} {\bibfnamefont {H.}~\bibnamefont {Weng}}, \bibinfo
  {author} {\bibfnamefont {Z.}~\bibnamefont {Wang}}, \bibinfo {author}
  {\bibfnamefont {X.}~\bibnamefont {Dai}}, \ and\ \bibinfo {author}
  {\bibfnamefont {Z.}~\bibnamefont {Fang}},\ }\href {\doibase
  10.1103/PhysRevLett.107.186806} {\bibfield  {journal} {\bibinfo  {journal}
  {Phys. Rev. Lett.}\ }\textbf {\bibinfo {volume} {107}},\ \bibinfo {pages}
  {186806} (\bibinfo {year} {2011})}\BibitemShut {NoStop}%
\bibitem [{\citenamefont {Young}\ \emph {et~al.}(2012)\citenamefont {Young},
  \citenamefont {Zaheer}, \citenamefont {Teo}, \citenamefont {Kane},
  \citenamefont {Mele},\ and\ \citenamefont {Rappe}}]{Kane12}%
  \BibitemOpen
  \bibfield  {author} {\bibinfo {author} {\bibfnamefont {S.~M.}\ \bibnamefont
  {Young}}, \bibinfo {author} {\bibfnamefont {S.}~\bibnamefont {Zaheer}},
  \bibinfo {author} {\bibfnamefont {J.~C.~Y.}\ \bibnamefont {Teo}}, \bibinfo
  {author} {\bibfnamefont {C.~L.}\ \bibnamefont {Kane}}, \bibinfo {author}
  {\bibfnamefont {E.~J.}\ \bibnamefont {Mele}}, \ and\ \bibinfo {author}
  {\bibfnamefont {A.~M.}\ \bibnamefont {Rappe}},\ }\href {\doibase
  10.1103/PhysRevLett.108.140405} {\bibfield  {journal} {\bibinfo  {journal}
  {Phys. Rev. Lett.}\ }\textbf {\bibinfo {volume} {108}},\ \bibinfo {pages}
  {140405} (\bibinfo {year} {2012})}\BibitemShut {NoStop}%
\bibitem [{\citenamefont {Wang}\ \emph {et~al.}(2012)\citenamefont {Wang},
  \citenamefont {Sun}, \citenamefont {Chen}, \citenamefont {Franchini},
  \citenamefont {Xu}, \citenamefont {Weng}, \citenamefont {Dai},\ and\
  \citenamefont {Fang}}]{Fang12}%
  \BibitemOpen
  \bibfield  {author} {\bibinfo {author} {\bibfnamefont {Z.}~\bibnamefont
  {Wang}}, \bibinfo {author} {\bibfnamefont {Y.}~\bibnamefont {Sun}}, \bibinfo
  {author} {\bibfnamefont {X.-Q.}\ \bibnamefont {Chen}}, \bibinfo {author}
  {\bibfnamefont {C.}~\bibnamefont {Franchini}}, \bibinfo {author}
  {\bibfnamefont {G.}~\bibnamefont {Xu}}, \bibinfo {author} {\bibfnamefont
  {H.}~\bibnamefont {Weng}}, \bibinfo {author} {\bibfnamefont {X.}~\bibnamefont
  {Dai}}, \ and\ \bibinfo {author} {\bibfnamefont {Z.}~\bibnamefont {Fang}},\
  }\href {\doibase 10.1103/PhysRevB.85.195320} {\bibfield  {journal} {\bibinfo
  {journal} {Phys. Rev. B}\ }\textbf {\bibinfo {volume} {85}},\ \bibinfo
  {pages} {195320} (\bibinfo {year} {2012})}\BibitemShut {NoStop}%
\bibitem [{\citenamefont {Wang}\ \emph {et~al.}(2013)\citenamefont {Wang},
  \citenamefont {Weng}, \citenamefont {Wu}, \citenamefont {Dai},\ and\
  \citenamefont {Fang}}]{Fang13}%
  \BibitemOpen
  \bibfield  {author} {\bibinfo {author} {\bibfnamefont {Z.}~\bibnamefont
  {Wang}}, \bibinfo {author} {\bibfnamefont {H.}~\bibnamefont {Weng}}, \bibinfo
  {author} {\bibfnamefont {Q.}~\bibnamefont {Wu}}, \bibinfo {author}
  {\bibfnamefont {X.}~\bibnamefont {Dai}}, \ and\ \bibinfo {author}
  {\bibfnamefont {Z.}~\bibnamefont {Fang}},\ }\href {\doibase
  10.1103/PhysRevB.88.125427} {\bibfield  {journal} {\bibinfo  {journal} {Phys.
  Rev. B}\ }\textbf {\bibinfo {volume} {88}},\ \bibinfo {pages} {125427}
  (\bibinfo {year} {2013})}\BibitemShut {NoStop}%
\bibitem [{\citenamefont {Liu}\ \emph {et~al.}(2014)\citenamefont {Liu},
  \citenamefont {Zhou}, \citenamefont {Zhang}, \citenamefont {Wang},
  \citenamefont {Weng}, \citenamefont {Prabhakaran}, \citenamefont {Mo},
  \citenamefont {Shen}, \citenamefont {Fang}, \citenamefont {Dai},
  \citenamefont {Hussain},\ and\ \citenamefont {Chen}}]{Chen14}%
  \BibitemOpen
  \bibfield  {author} {\bibinfo {author} {\bibfnamefont {Z.~K.}\ \bibnamefont
  {Liu}}, \bibinfo {author} {\bibfnamefont {B.}~\bibnamefont {Zhou}}, \bibinfo
  {author} {\bibfnamefont {Y.}~\bibnamefont {Zhang}}, \bibinfo {author}
  {\bibfnamefont {Z.~J.}\ \bibnamefont {Wang}}, \bibinfo {author}
  {\bibfnamefont {H.~M.}\ \bibnamefont {Weng}}, \bibinfo {author}
  {\bibfnamefont {D.}~\bibnamefont {Prabhakaran}}, \bibinfo {author}
  {\bibfnamefont {S.-K.}\ \bibnamefont {Mo}}, \bibinfo {author} {\bibfnamefont
  {Z.~X.}\ \bibnamefont {Shen}}, \bibinfo {author} {\bibfnamefont
  {Z.}~\bibnamefont {Fang}}, \bibinfo {author} {\bibfnamefont {X.}~\bibnamefont
  {Dai}}, \bibinfo {author} {\bibfnamefont {Z.}~\bibnamefont {Hussain}}, \ and\
  \bibinfo {author} {\bibfnamefont {Y.~L.}\ \bibnamefont {Chen}},\ }\href
  {\doibase 10.1126/science.1245085} {\bibfield  {journal} {\bibinfo  {journal}
  {Science}\ }\textbf {\bibinfo {volume} {343}},\ \bibinfo {pages} {864}
  (\bibinfo {year} {2014})}\BibitemShut {NoStop}%
\bibitem [{\citenamefont {Neupane}\ \emph {et~al.}(2014)\citenamefont
  {Neupane}, \citenamefont {Xu}, \citenamefont {Sankar}, \citenamefont
  {Alidoust}, \citenamefont {Bian}, \citenamefont {Liu}, \citenamefont
  {Belopolski}, \citenamefont {Chang}, \citenamefont {Jeng}, \citenamefont
  {Lin}, \citenamefont {Bansil}, \citenamefont {Chou},\ and\ \citenamefont
  {Hasan}}]{Neupane14}%
  \BibitemOpen
  \bibfield  {author} {\bibinfo {author} {\bibfnamefont {M.}~\bibnamefont
  {Neupane}}, \bibinfo {author} {\bibfnamefont {S.-Y.}\ \bibnamefont {Xu}},
  \bibinfo {author} {\bibfnamefont {R.}~\bibnamefont {Sankar}}, \bibinfo
  {author} {\bibfnamefont {N.}~\bibnamefont {Alidoust}}, \bibinfo {author}
  {\bibfnamefont {G.}~\bibnamefont {Bian}}, \bibinfo {author} {\bibfnamefont
  {C.}~\bibnamefont {Liu}}, \bibinfo {author} {\bibfnamefont {I.}~\bibnamefont
  {Belopolski}}, \bibinfo {author} {\bibfnamefont {T.-R.}\ \bibnamefont
  {Chang}}, \bibinfo {author} {\bibfnamefont {H.-T.}\ \bibnamefont {Jeng}},
  \bibinfo {author} {\bibfnamefont {H.}~\bibnamefont {Lin}}, \bibinfo {author}
  {\bibfnamefont {A.}~\bibnamefont {Bansil}}, \bibinfo {author} {\bibfnamefont
  {F.}~\bibnamefont {Chou}}, \ and\ \bibinfo {author} {\bibfnamefont {M.~Z.}\
  \bibnamefont {Hasan}},\ }\href {http://dx.doi.org/10.1038/ncomms4786}
  {\bibfield  {journal} {\bibinfo  {journal} {Nat. Commun.}\ }\textbf {\bibinfo
  {volume} {5}} (\bibinfo {year} {2014})}\BibitemShut {NoStop}%
\bibitem [{\citenamefont {Xu}\ \emph {et~al.}(2015)\citenamefont {Xu},
  \citenamefont {Belopolski}, \citenamefont {Alidoust}, \citenamefont
  {Neupane}, \citenamefont {Bian}, \citenamefont {Zhang}, \citenamefont
  {Sankar}, \citenamefont {Chang}, \citenamefont {Yuan}, \citenamefont {Lee},
  \citenamefont {Huang}, \citenamefont {Zheng}, \citenamefont {Ma},
  \citenamefont {Sanchez}, \citenamefont {Wang}, \citenamefont {Bansil},
  \citenamefont {Chou}, \citenamefont {Shibayev}, \citenamefont {Lin},
  \citenamefont {Jia},\ and\ \citenamefont {Hasan}}]{HasanTaAs}%
  \BibitemOpen
  \bibfield  {author} {\bibinfo {author} {\bibfnamefont {S.-Y.}\ \bibnamefont
  {Xu}}, \bibinfo {author} {\bibfnamefont {I.}~\bibnamefont {Belopolski}},
  \bibinfo {author} {\bibfnamefont {N.}~\bibnamefont {Alidoust}}, \bibinfo
  {author} {\bibfnamefont {M.}~\bibnamefont {Neupane}}, \bibinfo {author}
  {\bibfnamefont {G.}~\bibnamefont {Bian}}, \bibinfo {author} {\bibfnamefont
  {C.}~\bibnamefont {Zhang}}, \bibinfo {author} {\bibfnamefont
  {R.}~\bibnamefont {Sankar}}, \bibinfo {author} {\bibfnamefont
  {G.}~\bibnamefont {Chang}}, \bibinfo {author} {\bibfnamefont
  {Z.}~\bibnamefont {Yuan}}, \bibinfo {author} {\bibfnamefont {C.-C.}\
  \bibnamefont {Lee}}, \bibinfo {author} {\bibfnamefont {S.-M.}\ \bibnamefont
  {Huang}}, \bibinfo {author} {\bibfnamefont {H.}~\bibnamefont {Zheng}},
  \bibinfo {author} {\bibfnamefont {J.}~\bibnamefont {Ma}}, \bibinfo {author}
  {\bibfnamefont {D.~S.}\ \bibnamefont {Sanchez}}, \bibinfo {author}
  {\bibfnamefont {B.}~\bibnamefont {Wang}}, \bibinfo {author} {\bibfnamefont
  {A.}~\bibnamefont {Bansil}}, \bibinfo {author} {\bibfnamefont
  {F.}~\bibnamefont {Chou}}, \bibinfo {author} {\bibfnamefont {P.~P.}\
  \bibnamefont {Shibayev}}, \bibinfo {author} {\bibfnamefont {H.}~\bibnamefont
  {Lin}}, \bibinfo {author} {\bibfnamefont {S.}~\bibnamefont {Jia}}, \ and\
  \bibinfo {author} {\bibfnamefont {M.~Z.}\ \bibnamefont {Hasan}},\ }\href
  {\doibase 10.1126/science.aaa9297} {\bibfield  {journal} {\bibinfo  {journal}
  {Science}\ }\textbf {\bibinfo {volume} {349}},\ \bibinfo {pages} {613}
  (\bibinfo {year} {2015})}\BibitemShut {NoStop}%
\bibitem [{\citenamefont {Lv}\ \emph {et~al.}(2015{\natexlab{a}})\citenamefont
  {Lv}, \citenamefont {Xu}, \citenamefont {Weng}, \citenamefont {Ma},
  \citenamefont {Richard}, \citenamefont {Huang}, \citenamefont {Zhao},
  \citenamefont {Chen}, \citenamefont {Matt}, \citenamefont {Bisti},
  \citenamefont {Strocov}, \citenamefont {Mesot}, \citenamefont {Fang},
  \citenamefont {Dai}, \citenamefont {Qian}, \citenamefont {Shi},\ and\
  \citenamefont {Ding}}]{DingTaAs2}%
  \BibitemOpen
  \bibfield  {author} {\bibinfo {author} {\bibfnamefont {B.~Q.}\ \bibnamefont
  {Lv}}, \bibinfo {author} {\bibfnamefont {N.}~\bibnamefont {Xu}}, \bibinfo
  {author} {\bibfnamefont {H.~M.}\ \bibnamefont {Weng}}, \bibinfo {author}
  {\bibfnamefont {J.~Z.}\ \bibnamefont {Ma}}, \bibinfo {author} {\bibfnamefont
  {P.}~\bibnamefont {Richard}}, \bibinfo {author} {\bibfnamefont {X.~C.}\
  \bibnamefont {Huang}}, \bibinfo {author} {\bibfnamefont {L.~X.}\ \bibnamefont
  {Zhao}}, \bibinfo {author} {\bibfnamefont {G.~F.}\ \bibnamefont {Chen}},
  \bibinfo {author} {\bibfnamefont {C.~E.}\ \bibnamefont {Matt}}, \bibinfo
  {author} {\bibfnamefont {F.}~\bibnamefont {Bisti}}, \bibinfo {author}
  {\bibfnamefont {V.~N.}\ \bibnamefont {Strocov}}, \bibinfo {author}
  {\bibfnamefont {J.}~\bibnamefont {Mesot}}, \bibinfo {author} {\bibfnamefont
  {Z.}~\bibnamefont {Fang}}, \bibinfo {author} {\bibfnamefont {X.}~\bibnamefont
  {Dai}}, \bibinfo {author} {\bibfnamefont {T.}~\bibnamefont {Qian}}, \bibinfo
  {author} {\bibfnamefont {M.}~\bibnamefont {Shi}}, \ and\ \bibinfo {author}
  {\bibfnamefont {H.}~\bibnamefont {Ding}},\ }\href
  {http://dx.doi.org/10.1038/nphys3426} {\bibfield  {journal} {\bibinfo
  {journal} {Nat Phys}\ }\textbf {\bibinfo {volume} {11}},\ \bibinfo {pages}
  {724} (\bibinfo {year} {2015}{\natexlab{a}})}\BibitemShut {NoStop}%
\bibitem [{\citenamefont {Lv}\ \emph {et~al.}(2015{\natexlab{b}})\citenamefont
  {Lv}, \citenamefont {Weng}, \citenamefont {Fu}, \citenamefont {Wang},
  \citenamefont {Miao}, \citenamefont {Ma}, \citenamefont {Richard},
  \citenamefont {Huang}, \citenamefont {Zhao}, \citenamefont {Chen},
  \citenamefont {Fang}, \citenamefont {Dai}, \citenamefont {Qian},\ and\
  \citenamefont {Ding}}]{DingTaAs}%
  \BibitemOpen
  \bibfield  {author} {\bibinfo {author} {\bibfnamefont {B.~Q.}\ \bibnamefont
  {Lv}}, \bibinfo {author} {\bibfnamefont {H.~M.}\ \bibnamefont {Weng}},
  \bibinfo {author} {\bibfnamefont {B.~B.}\ \bibnamefont {Fu}}, \bibinfo
  {author} {\bibfnamefont {X.~P.}\ \bibnamefont {Wang}}, \bibinfo {author}
  {\bibfnamefont {H.}~\bibnamefont {Miao}}, \bibinfo {author} {\bibfnamefont
  {J.}~\bibnamefont {Ma}}, \bibinfo {author} {\bibfnamefont {P.}~\bibnamefont
  {Richard}}, \bibinfo {author} {\bibfnamefont {X.~C.}\ \bibnamefont {Huang}},
  \bibinfo {author} {\bibfnamefont {L.~X.}\ \bibnamefont {Zhao}}, \bibinfo
  {author} {\bibfnamefont {G.~F.}\ \bibnamefont {Chen}}, \bibinfo {author}
  {\bibfnamefont {Z.}~\bibnamefont {Fang}}, \bibinfo {author} {\bibfnamefont
  {X.}~\bibnamefont {Dai}}, \bibinfo {author} {\bibfnamefont {T.}~\bibnamefont
  {Qian}}, \ and\ \bibinfo {author} {\bibfnamefont {H.}~\bibnamefont {Ding}},\
  }\href {\doibase 10.1103/PhysRevX.5.031013} {\bibfield  {journal} {\bibinfo
  {journal} {Phys. Rev. X}\ }\textbf {\bibinfo {volume} {5}},\ \bibinfo {pages}
  {031013} (\bibinfo {year} {2015}{\natexlab{b}})}\BibitemShut {NoStop}%
\bibitem [{\citenamefont {Lu}\ \emph {et~al.}(2015)\citenamefont {Lu},
  \citenamefont {Wang}, \citenamefont {Ye}, \citenamefont {Ran}, \citenamefont
  {Fu}, \citenamefont {Joannopoulos},\ and\ \citenamefont {Solja{\v
  c}i{\'c}}}]{Lu15}%
  \BibitemOpen
  \bibfield  {author} {\bibinfo {author} {\bibfnamefont {L.}~\bibnamefont
  {Lu}}, \bibinfo {author} {\bibfnamefont {Z.}~\bibnamefont {Wang}}, \bibinfo
  {author} {\bibfnamefont {D.}~\bibnamefont {Ye}}, \bibinfo {author}
  {\bibfnamefont {L.}~\bibnamefont {Ran}}, \bibinfo {author} {\bibfnamefont
  {L.}~\bibnamefont {Fu}}, \bibinfo {author} {\bibfnamefont {J.~D.}\
  \bibnamefont {Joannopoulos}}, \ and\ \bibinfo {author} {\bibfnamefont
  {M.}~\bibnamefont {Solja{\v c}i{\'c}}},\ }\href {\doibase
  10.1126/science.aaa9273} {\bibfield  {journal} {\bibinfo  {journal}
  {Science}\ }\textbf {\bibinfo {volume} {349}},\ \bibinfo {pages} {622}
  (\bibinfo {year} {2015})}\BibitemShut {NoStop}%
\bibitem [{\citenamefont {Adler}(1969)}]{Adler69}%
  \BibitemOpen
  \bibfield  {author} {\bibinfo {author} {\bibfnamefont {S.~L.}\ \bibnamefont
  {Adler}},\ }\href {\doibase 10.1103/PhysRev.177.2426} {\bibfield  {journal}
  {\bibinfo  {journal} {Phys. Rev.}\ }\textbf {\bibinfo {volume} {177}},\
  \bibinfo {pages} {2426} (\bibinfo {year} {1969})}\BibitemShut {NoStop}%
\bibitem [{\citenamefont {Bell}\ and\ \citenamefont {Jackiw}(1969)}]{Jackiw69}%
  \BibitemOpen
  \bibfield  {author} {\bibinfo {author} {\bibfnamefont {J.~S.}\ \bibnamefont
  {Bell}}\ and\ \bibinfo {author} {\bibfnamefont {R.}~\bibnamefont {Jackiw}},\
  }\href@noop {} {\bibfield  {journal} {\bibinfo  {journal} {Nuovo Cimento A}\
  }\textbf {\bibinfo {volume} {60}},\ \bibinfo {pages} {4} (\bibinfo {year}
  {1969})}\BibitemShut {NoStop}%
\bibitem [{\citenamefont {Xiong}\ \emph {et~al.}(2015)\citenamefont {Xiong},
  \citenamefont {Kushwaha}, \citenamefont {Liang}, \citenamefont {Krizan},
  \citenamefont {Hirschberger}, \citenamefont {Wang}, \citenamefont {Cava},\
  and\ \citenamefont {Ong}}]{Ong_anomaly}%
  \BibitemOpen
  \bibfield  {author} {\bibinfo {author} {\bibfnamefont {J.}~\bibnamefont
  {Xiong}}, \bibinfo {author} {\bibfnamefont {S.~K.}\ \bibnamefont {Kushwaha}},
  \bibinfo {author} {\bibfnamefont {T.}~\bibnamefont {Liang}}, \bibinfo
  {author} {\bibfnamefont {J.~W.}\ \bibnamefont {Krizan}}, \bibinfo {author}
  {\bibfnamefont {M.}~\bibnamefont {Hirschberger}}, \bibinfo {author}
  {\bibfnamefont {W.}~\bibnamefont {Wang}}, \bibinfo {author} {\bibfnamefont
  {R.~J.}\ \bibnamefont {Cava}}, \ and\ \bibinfo {author} {\bibfnamefont
  {N.~P.}\ \bibnamefont {Ong}},\ }\href {\doibase 10.1126/science.aac6089}
  {\bibfield  {journal} {\bibinfo  {journal} {Science}\ }\textbf {\bibinfo
  {volume} {350}},\ \bibinfo {pages} {413} (\bibinfo {year}
  {2015})}\BibitemShut {NoStop}%
\bibitem [{\citenamefont {Li}\ \emph {et~al.}(2016)\citenamefont {Li},
  \citenamefont {Kharzeev}, \citenamefont {Zhang}, \citenamefont {Huang},
  \citenamefont {Pletikosic}, \citenamefont {Fedorov}, \citenamefont {Zhong},
  \citenamefont {Schneeloch}, \citenamefont {Gu},\ and\ \citenamefont
  {Valla}}]{Li_anomaly}%
  \BibitemOpen
  \bibfield  {author} {\bibinfo {author} {\bibfnamefont {Q.}~\bibnamefont
  {Li}}, \bibinfo {author} {\bibfnamefont {D.~E.}\ \bibnamefont {Kharzeev}},
  \bibinfo {author} {\bibfnamefont {C.}~\bibnamefont {Zhang}}, \bibinfo
  {author} {\bibfnamefont {Y.}~\bibnamefont {Huang}}, \bibinfo {author}
  {\bibfnamefont {I.}~\bibnamefont {Pletikosic}}, \bibinfo {author}
  {\bibfnamefont {A.~V.}\ \bibnamefont {Fedorov}}, \bibinfo {author}
  {\bibfnamefont {R.~D.}\ \bibnamefont {Zhong}}, \bibinfo {author}
  {\bibfnamefont {J.~A.}\ \bibnamefont {Schneeloch}}, \bibinfo {author}
  {\bibfnamefont {G.~D.}\ \bibnamefont {Gu}}, \ and\ \bibinfo {author}
  {\bibfnamefont {T.}~\bibnamefont {Valla}},\ }\href
  {http://dx.doi.org/10.1038/nphys3648} {\bibfield  {journal} {\bibinfo
  {journal} {Nat Phys}\ }\textbf {\bibinfo {volume} {12}},\ \bibinfo {pages}
  {550} (\bibinfo {year} {2016})}\BibitemShut {NoStop}%
\bibitem [{\citenamefont {Hirschberger}\ \emph {et~al.}(2016)\citenamefont
  {Hirschberger}, \citenamefont {Kushwaha}, \citenamefont {Wang}, \citenamefont
  {Gibson}, \citenamefont {Liang}, \citenamefont {Belvin}, \citenamefont
  {Bernevig}, \citenamefont {Cava},\ and\ \citenamefont {Ong}}]{Ong_GdPtBi}%
  \BibitemOpen
  \bibfield  {author} {\bibinfo {author} {\bibfnamefont {M.}~\bibnamefont
  {Hirschberger}}, \bibinfo {author} {\bibfnamefont {S.}~\bibnamefont
  {Kushwaha}}, \bibinfo {author} {\bibfnamefont {Z.}~\bibnamefont {Wang}},
  \bibinfo {author} {\bibfnamefont {Q.}~\bibnamefont {Gibson}}, \bibinfo
  {author} {\bibfnamefont {S.}~\bibnamefont {Liang}}, \bibinfo {author}
  {\bibfnamefont {C.~A.}\ \bibnamefont {Belvin}}, \bibinfo {author}
  {\bibfnamefont {B.~A.}\ \bibnamefont {Bernevig}}, \bibinfo {author}
  {\bibfnamefont {R.~J.}\ \bibnamefont {Cava}}, \ and\ \bibinfo {author}
  {\bibfnamefont {N.~P.}\ \bibnamefont {Ong}},\ }\href
  {http://dx.doi.org/10.1038/nmat4684} {\bibfield  {journal} {\bibinfo
  {journal} {Nat Mater}\ }\textbf {\bibinfo {volume} {15}},\ \bibinfo {pages}
  {1161} (\bibinfo {year} {2016})}\BibitemShut {NoStop}%
\bibitem [{\citenamefont {Son}\ and\ \citenamefont {Spivak}(2013)}]{Spivak12}%
  \BibitemOpen
  \bibfield  {author} {\bibinfo {author} {\bibfnamefont {D.~T.}\ \bibnamefont
  {Son}}\ and\ \bibinfo {author} {\bibfnamefont {B.~Z.}\ \bibnamefont
  {Spivak}},\ }\href {\doibase 10.1103/PhysRevB.88.104412} {\bibfield
  {journal} {\bibinfo  {journal} {Phys. Rev. B}\ }\textbf {\bibinfo {volume}
  {88}},\ \bibinfo {pages} {104412} (\bibinfo {year} {2013})}\BibitemShut
  {NoStop}%
\bibitem [{\citenamefont {Burkov}(2014)}]{Burkov_lmr_prl}%
  \BibitemOpen
  \bibfield  {author} {\bibinfo {author} {\bibfnamefont {A.~A.}\ \bibnamefont
  {Burkov}},\ }\href {\doibase 10.1103/PhysRevLett.113.247203} {\bibfield
  {journal} {\bibinfo  {journal} {Phys. Rev. Lett.}\ }\textbf {\bibinfo
  {volume} {113}},\ \bibinfo {pages} {247203} (\bibinfo {year}
  {2014})}\BibitemShut {NoStop}%
\bibitem [{\citenamefont {Burkov}(2015)}]{Burkov_lmr_prb}%
  \BibitemOpen
  \bibfield  {author} {\bibinfo {author} {\bibfnamefont {A.~A.}\ \bibnamefont
  {Burkov}},\ }\href {\doibase 10.1103/PhysRevB.91.245157} {\bibfield
  {journal} {\bibinfo  {journal} {Phys. Rev. B}\ }\textbf {\bibinfo {volume}
  {91}},\ \bibinfo {pages} {245157} (\bibinfo {year} {2015})}\BibitemShut
  {NoStop}%
\bibitem [{\citenamefont {Burkov}\ and\ \citenamefont {Kim}(2016)}]{Burkov16}%
  \BibitemOpen
  \bibfield  {author} {\bibinfo {author} {\bibfnamefont {A.~A.}\ \bibnamefont
  {Burkov}}\ and\ \bibinfo {author} {\bibfnamefont {Y.~B.}\ \bibnamefont
  {Kim}},\ }\href {\doibase 10.1103/PhysRevLett.117.136602} {\bibfield
  {journal} {\bibinfo  {journal} {Phys. Rev. Lett.}\ }\textbf {\bibinfo
  {volume} {117}},\ \bibinfo {pages} {136602} (\bibinfo {year}
  {2016})}\BibitemShut {NoStop}%
\bibitem [{\citenamefont {Parameswaran}\ \emph {et~al.}(2014)\citenamefont
  {Parameswaran}, \citenamefont {Grover}, \citenamefont {Abanin}, \citenamefont
  {Pesin},\ and\ \citenamefont {Vishwanath}}]{Parameswaran14}%
  \BibitemOpen
  \bibfield  {author} {\bibinfo {author} {\bibfnamefont {S.~A.}\ \bibnamefont
  {Parameswaran}}, \bibinfo {author} {\bibfnamefont {T.}~\bibnamefont
  {Grover}}, \bibinfo {author} {\bibfnamefont {D.~A.}\ \bibnamefont {Abanin}},
  \bibinfo {author} {\bibfnamefont {D.~A.}\ \bibnamefont {Pesin}}, \ and\
  \bibinfo {author} {\bibfnamefont {A.}~\bibnamefont {Vishwanath}},\ }\href
  {\doibase 10.1103/PhysRevX.4.031035} {\bibfield  {journal} {\bibinfo
  {journal} {Phys. Rev. X}\ }\textbf {\bibinfo {volume} {4}},\ \bibinfo {pages}
  {031035} (\bibinfo {year} {2014})}\BibitemShut {NoStop}%
\bibitem [{\citenamefont {Zyuzin}\ \emph {et~al.}(2012)\citenamefont {Zyuzin},
  \citenamefont {Wu},\ and\ \citenamefont {Burkov}}]{Zyuzin12-2}%
  \BibitemOpen
  \bibfield  {author} {\bibinfo {author} {\bibfnamefont {A.~A.}\ \bibnamefont
  {Zyuzin}}, \bibinfo {author} {\bibfnamefont {S.}~\bibnamefont {Wu}}, \ and\
  \bibinfo {author} {\bibfnamefont {A.~A.}\ \bibnamefont {Burkov}},\ }\href
  {\doibase 10.1103/PhysRevB.85.165110} {\bibfield  {journal} {\bibinfo
  {journal} {Phys. Rev. B}\ }\textbf {\bibinfo {volume} {85}},\ \bibinfo
  {pages} {165110} (\bibinfo {year} {2012})}\BibitemShut {NoStop}%
\bibitem [{\citenamefont {Fukushima}\ \emph {et~al.}(2008)\citenamefont
  {Fukushima}, \citenamefont {Kharzeev},\ and\ \citenamefont
  {Warringa}}]{Kharzeev08}%
  \BibitemOpen
  \bibfield  {author} {\bibinfo {author} {\bibfnamefont {K.}~\bibnamefont
  {Fukushima}}, \bibinfo {author} {\bibfnamefont {D.~E.}\ \bibnamefont
  {Kharzeev}}, \ and\ \bibinfo {author} {\bibfnamefont {H.~J.}\ \bibnamefont
  {Warringa}},\ }\href {\doibase 10.1103/PhysRevD.78.074033} {\bibfield
  {journal} {\bibinfo  {journal} {Phys. Rev. D}\ }\textbf {\bibinfo {volume}
  {78}},\ \bibinfo {pages} {074033} (\bibinfo {year} {2008})}\BibitemShut
  {NoStop}%
\bibitem [{\citenamefont {Vazifeh}\ and\ \citenamefont
  {Franz}(2013)}]{Franz13}%
  \BibitemOpen
  \bibfield  {author} {\bibinfo {author} {\bibfnamefont {M.~M.}\ \bibnamefont
  {Vazifeh}}\ and\ \bibinfo {author} {\bibfnamefont {M.}~\bibnamefont
  {Franz}},\ }\href {\doibase 10.1103/PhysRevLett.111.027201} {\bibfield
  {journal} {\bibinfo  {journal} {Phys. Rev. Lett.}\ }\textbf {\bibinfo
  {volume} {111}},\ \bibinfo {pages} {027201} (\bibinfo {year}
  {2013})}\BibitemShut {NoStop}%
\bibitem [{\citenamefont {Chen}\ \emph {et~al.}(2013)\citenamefont {Chen},
  \citenamefont {Wu},\ and\ \citenamefont {Burkov}}]{Chen13}%
  \BibitemOpen
  \bibfield  {author} {\bibinfo {author} {\bibfnamefont {Y.}~\bibnamefont
  {Chen}}, \bibinfo {author} {\bibfnamefont {S.}~\bibnamefont {Wu}}, \ and\
  \bibinfo {author} {\bibfnamefont {A.~A.}\ \bibnamefont {Burkov}},\ }\href
  {\doibase 10.1103/PhysRevB.88.125105} {\bibfield  {journal} {\bibinfo
  {journal} {Phys. Rev. B}\ }\textbf {\bibinfo {volume} {88}},\ \bibinfo
  {pages} {125105} (\bibinfo {year} {2013})}\BibitemShut {NoStop}%
\bibitem [{\citenamefont {Altland}\ and\ \citenamefont
  {Bagrets}(2016)}]{Altland15}%
  \BibitemOpen
  \bibfield  {author} {\bibinfo {author} {\bibfnamefont {A.}~\bibnamefont
  {Altland}}\ and\ \bibinfo {author} {\bibfnamefont {D.}~\bibnamefont
  {Bagrets}},\ }\href {\doibase 10.1103/PhysRevB.93.075113} {\bibfield
  {journal} {\bibinfo  {journal} {Phys. Rev. B}\ }\textbf {\bibinfo {volume}
  {93}},\ \bibinfo {pages} {075113} (\bibinfo {year} {2016})}\BibitemShut
  {NoStop}%
\bibitem [{\citenamefont {Pan}(1957)}]{Pan57}%
  \BibitemOpen
  \bibfield  {author} {\bibinfo {author} {\bibfnamefont {J.~P.}\ \bibnamefont
  {Pan}},\ }\href@noop {} {\emph {\bibinfo {title} {Solid State Physics}}},\
  edited by\ \bibinfo {editor} {\bibfnamefont {F.}~\bibnamefont {Seitz}}\ and\
  \bibinfo {editor} {\bibfnamefont {D.}~\bibnamefont {Turnbull}},\
  Vol.~\bibinfo {volume} {5}\ (\bibinfo  {publisher} {Academic, New York},\
  \bibinfo {year} {1957})\ pp.\ \bibinfo {pages} {1--96}\BibitemShut {NoStop}%
\bibitem [{\citenamefont {Hong}\ and\ \citenamefont
  {Giordano}(1995)}]{Giordano95}%
  \BibitemOpen
  \bibfield  {author} {\bibinfo {author} {\bibfnamefont {K.}~\bibnamefont
  {Hong}}\ and\ \bibinfo {author} {\bibfnamefont {N.}~\bibnamefont
  {Giordano}},\ }\href {\doibase 10.1103/PhysRevB.51.9855} {\bibfield
  {journal} {\bibinfo  {journal} {Phys. Rev. B}\ }\textbf {\bibinfo {volume}
  {51}},\ \bibinfo {pages} {9855} (\bibinfo {year} {1995})}\BibitemShut
  {NoStop}%
\bibitem [{\citenamefont {Tang}\ \emph {et~al.}(2003)\citenamefont {Tang},
  \citenamefont {Kawakami}, \citenamefont {Awschalom},\ and\ \citenamefont
  {Roukes}}]{Awschalom_gphe}%
  \BibitemOpen
  \bibfield  {author} {\bibinfo {author} {\bibfnamefont {H.~X.}\ \bibnamefont
  {Tang}}, \bibinfo {author} {\bibfnamefont {R.~K.}\ \bibnamefont {Kawakami}},
  \bibinfo {author} {\bibfnamefont {D.~D.}\ \bibnamefont {Awschalom}}, \ and\
  \bibinfo {author} {\bibfnamefont {M.~L.}\ \bibnamefont {Roukes}},\ }\href
  {\doibase 10.1103/PhysRevLett.90.107201} {\bibfield  {journal} {\bibinfo
  {journal} {Phys. Rev. Lett.}\ }\textbf {\bibinfo {volume} {90}},\ \bibinfo
  {pages} {107201} (\bibinfo {year} {2003})}\BibitemShut {NoStop}%
\bibitem [{\citenamefont {{Liang}}\ \emph {et~al.}(2016)\citenamefont
  {{Liang}}, \citenamefont {{Gibson}}, \citenamefont {{Liu}}, \citenamefont
  {{Wang}}, \citenamefont {{Cava}},\ and\ \citenamefont {{Ong}}}]{Ong16}%
  \BibitemOpen
  \bibfield  {author} {\bibinfo {author} {\bibfnamefont {T.}~\bibnamefont
  {{Liang}}}, \bibinfo {author} {\bibfnamefont {Q.}~\bibnamefont {{Gibson}}},
  \bibinfo {author} {\bibfnamefont {M.}~\bibnamefont {{Liu}}}, \bibinfo
  {author} {\bibfnamefont {W.}~\bibnamefont {{Wang}}}, \bibinfo {author}
  {\bibfnamefont {R.~J.}\ \bibnamefont {{Cava}}}, \ and\ \bibinfo {author}
  {\bibfnamefont {N.~P.}\ \bibnamefont {{Ong}}},\ }\href@noop {} {\bibfield
  {journal} {\bibinfo  {journal} {ArXiv e-prints}\ } (\bibinfo {year}
  {2016})},\ \Eprint {http://arxiv.org/abs/1612.06972} {arXiv:1612.06972
  [cond-mat.mes-hall]} \BibitemShut {NoStop}%
\end{thebibliography}%

\end{document}